# Attractor and integrator networks in the brain

Mikail Khona[1,2] and Ila R. Fiete [*,1]

[1]Department of Brain and Cognitive Sciences & McGovern Institute, MIT
[2]Department of Physics, MIT

## Abstract

In this review, we describe the singular success of attractor neural network models in describing how the brain maintains persistent activity states for working memory, error-corrects, and integrates noisy cues. We consider the mechanisms by which simple and forgetful units can organize to collectively generate dynamics on the long time-scales required for such computations. We discuss the myriad potential uses of attractor dynamics for computation in the brain, and showcase notable examples of brain systems in which inherently low-dimensional continuous attractor dynamics have been concretely and rigorously identified. Thus, it is now possible to conclusively state that the brain constructs and uses such systems for computation. Finally, we look ahead by highlighting recent theoretical advances in understanding how the fundamental tradeoffs between robustness and capacity and between structure and flexibility can be overcome by reusing and recombining the same set of modular attractors for multiple functions, so they together produce representations that are structurally constrained and robust but exhibit high capacity and are flexible.

## Introduction

One of Biology's grand challenges is to explain how order and complex function spring from inanimate physical systems composed of much simpler parts. The brain creates order in its representations of the world and performs complex functions through the collective interactions of simpler elements. In this review, we will describe and evaluate the hypothesis that attractor dynamics in widespread regions of the central nervous system play a key role in constructing some of these representations, generating long time-scales to support integration and memory functions, and endowing all these functions with robustness. We will review the specific predictions of attractor-based models and the now-extensive body of work testing these predictions. Thus, we will illustrate that the theory and validation of computation with attractor dynamics in the brain is one of the biggest success stories in systems neuroscience.

Some of the first formal circuit-level models of brain function focused on the problem of associative memory and how neural circuits might generate spatially distributed, stable patterns of activity that could function as such a memory [1, 2, 3, 4]. Hopfield networks, with multiple stable states learned from distributed input patterns, were proposed over four decades ago [5, 3, 6]. Network models possessing a continuous set of stable states, that could be used to represent continuous variables, were also first proposed around the same period [7]. Subsequently, many canonical brain circuits for motor control, sensory amplification and memory, motion integration, evidence integration, decision making, and spatial navigation have been modeled using the same general principle – that a set of states stabilized through collective positive feedback can be used for robust representation, memory, and to perform computations that involve memory [8, 9, 10, 11, 12, 13, 14, 15, 16, 17].

---

[*]Corresponding author

Because these are circuit-level models, but were typically inspired by experimental characterization of neurons recorded singly or a few at a time, the patterns of connectivity and cell activity correlations in the models automatically became novel and relatively specific predictions about the population dynamics and architecture of such circuits. As we will discuss below, the combination of these prediction-rich (and often conceptually simple) models, modern experimental breakthroughs in the acquisition of cellular-resolution population activity data, and novel and rigorous analyses of such data based on the model predictions has led to the accumulation of a preponderance of evidence that the brain constructs and exploits attractor networks for performing several essential computations.

We will begin by defining attractors, then describe proposed mechanisms for the construction of attractor network models in neuroscience. We will provide an overview of why attractor networks can be important for computation in the brain and give criteria for determining whether a system has non-trivial attractor dynamics. After this groundwork we will discuss several examples of brain circuits with non-trivial attractor dynamics. Finally, we will end with new directions in our understanding of how these simple circuits could contribute to flexible computation through reuse in multiple contexts.

## What are attractors?

To define an attractor, we first define a dynamical system and its states. A *dynamical system* is a set of variables together with all the rules that determine their time-evolution. The instantaneous value of these variables is called the *state* of the system at that moment. The state is a point (vector) in the *state space* of the dynamical system. An *attractor* is the minimal set of states in a state space, to which all nearby states eventually flow [18]. One simple example of an attractor is a stable fixed point: all neighboring states flow to it. Porting these crisp mathematical definitions to the brain involves challenges and simplifications, which revolve around identifying a sufficiently self-contained system and the variables necessary to determine its dynamics.

**Defining the state of a neural system:** Inherent in the definition of a dynamical system is the assumption that there are no external dynamical inputs to the system (equivalently, the system definition includes all such variables). The first simplification in characterizing the dynamics of a neural circuit is to assume that at least on the time-scale of interest, the system evolves *autonomously*. Given that subcircuits in the brain are interconnected with others, and that the brain itself interacts with the world, it is impossible to completely isolate these circuits into autonomous systems. However, we may define a notion of *effectively autonomous* dynamics over time-scales where inputs are not temporally varying and are untuned in the sense that they do not provide differential drive to subsets of the putative set of attractor states. The second simplification is in defining the states of the system. The time-evolution of a circuit in the brain may depend on the detailed pattern of all the spikes in all neurons, the levels of associated ions, neurotransmitters and modulators, the states of the ion channels. The weights and connections between neurons may be considered as parameters rather than variables on short timescales, but as variables if considering a longer time-scale. One widely used simplification in describing a neural circuit on the timescale of seconds is to use just the spiking outputs of the neurons in the circuit as the states, often further simplified as time-varying spike rates. If such a description is sufficient to predict the evolution of the system at the relevant time-scales, it can be viewed as a reasonable dynamical system model of the circuit. Even though spike or spike rate descriptions ignore sub-cellular and molecular variables to make the grossly simplifying assumption that the relevant circuit dynamics are governed by spikes, the state space of a vertebrate microcircuit described in this way is nevertheless very high-dimensional, comprising the number of neurons in the circuit, or $10^2-10^7$ cells. As we will see below, such simplified models can nevertheless yield rich and accurate predictions about neural circuits.

Attractors exist in various flavors: an attractor may consist of a single state or a set of states that trace out a complex shape, such as a curved manifold [1], Fig. 1 (rightmost column). States on an attractor may be stationary, or might flow along the attractor to trace out trajectories that are periodic (limit cycles, Fig. 1f, rightmost column) or chaotic (dynamics that are inherently unpredictable due to high sensitivity to small changes in the state [21]). Various combinations of such attractors, of different dimension, geometry, and topology, may coexist in different regions of the state space of a single dynamical system. Typically, the set of attractors in a dynamical system comprises a small subset of the state space, and attractor manifolds are usually much lower-dimensional than the state space. In cases where a system has multiple attractor states, the initial condition determines the attractor state to which the system flows.

**Defining attractors in the presence of noise:** Any real physical system unavoidably behaves non-deterministically from the perspective of a model of the system. This is because one cannot observe and describe all variables, and all uncharacterized variables together with true stochastic sources of variation (e.g., synaptic signalling noise from stochastic vesicle release [22], ionic number fluctuations in processes like spike initiation [23], calcium signaling, fluctuations in small copy numbers of proteins [24]) serve as effective sources of noise in the model. Noise can buffet states so they do not strictly localize to the attractor described in a noise-free version of the model, and can drive the system to escape from an attractor over time. However, the general idea of attracting states remains in the following sense: If the system is initialized near such a state, it tends to flow toward it and subsequently remains localized around it, for extended periods. In sum, since attractor states are where systems tend to localize (when not externally driven), they should be observable in the autonomous dynamics of real systems. This basic property is the basis for the most fundamental and robust tests of attractor dynamics in neural systems, as we will discuss below. In a nutshell, the critical signatures of attractors in real systems (discussed in more detail in later sections of this review) can be summarized as: localization of the states of a system to a lower-dimensional subset, flow of the states towards the subset after perturbation, and long-time and (effectively) autonomous stability of states in that subset.

---

[1]**Attractor manifolds:** If the number of attractor states in a network is large and the points are close to one another, they can behave effectively as a continuous set. If this near-continuous set traces out a surface that is locally Euclidean, it is called a manifold. Nonlinear continuous attractor manifolds can be curved and topologically complex (e.g. rings, torii, etc., Fig. 1c-d, rightmost column [19, 20]).

# Mechanisms: The construction of neural attractors

The general principle underlying the formation of non-trivial attractor states in neural circuits is strong recurrent positive feedback [2]. Positive feedback fights activity decay to stabilize certain states, and has been conjectured by James, Hebb and others [25, 26, 27, 28] as the basis for the stabilization of memory traces and persistent activity in the brain. Which states become stabilized into attractors depends on how the network sculpts the positive feedback, which according to the synaptic hypothesis is determined by the synaptic weights.[29, 30, 31]

In general, characterizing the relationship between structure and function in a large collection of interacting elements is extremely difficult, as described by Anderson in "More is different" [32]. For instance, a large collection of simple polar 3-atom molecules of hydrogen and oxygen give rise to emergent phenomena like liquidness and wetness and freezing into a solid, which cannot be predicted through intuition or drawing box-and-arrow diagrams. On the flip side, there is also emergent simplicity, in that the transitions and properties of the emergent states can be described with very few key parameters and variables.

One way to characterize the relationship between synaptic weights and attractor dynamics is to ask what attractor states a given set of weights produces (the "forward" problem). With the given weights, one can simulate the circuit and explore the resulting dynamics to find attractors of the system. A more powerful method, the Lyapunov function approach, holds for symmetric weight matrices ($W_{ij} = W_{ji}$) and rate-based neural dynamics. For this class of models, a generalized energy function (the Lyapunov function), which is a function of the weights and neural activation function [6, 5, 2], analytically specifies the network's dynamics. Stable (unstable) attractor states are the energy minima (maxima) of the derived landscape, and the network's state flows downhill towards the attractors Fig. 2e in the way a ball rolls down a gravitational potential.

Another way to characterize the relationship between attractors and network structure is to consider the "inverse" problem: given a set of attractors, what network structure could generate it? Neuroscientists want to solve the inverse problem to make predictions about underlying mechanism (and since neural activations are more readily observed than synaptic weights, the inverse problem is more frequently encountered than the forward one), while evolution, the brain, and artificially intelligent systems need to solve the inverse problem to be able to perform computations that require a given type of attractor dynamics (which we will discuss below). Theoretical neuroscience has discovered some solutions to the inverse problem for different types of attractors, as we describe next.

**Discrete attractors.** A well-known prescription for creating a discrete set of stable attractors at user-defined points in state space is given by the Hopfield network model [5], Fig. 1a: an externally induced and distributed pattern of neural activity is encoded into the weights by a Hebbian-like learning rule that causes co-activated neurons to excite each other and inhibit all the rest. As a result, these patterns become stable attractor states. If a sufficiently small number of patterns are inscribed into the weights, they can be retrieved from partial or corrupted versions of the stored states, thus the network is a content-addressable memory. More generally, the attractors of simple rate-based networks without synaptic delays and with arbitrary symmetric weight matrices[3] consist entirely of fixed points. Some non-symmetric networks can also support point attractors [33], but this is not the generic case and can require additional mechanisms like

---

[2] Non-trivial attractor states refer to any state other than the null activity state. Positive feedback is not synonymous with excitatory feedback: like mutual excitation, *disinhibition* or inhibition of one's inhibitor is also a form of positive feedback.

[3] A symmetric weight matrix $W$ satisfies $W^T = W$: it is invariant to reflection of its entries about its diagonal.

homeostatic plasticity [34, 35].

The attractor states in Hopfield-like networks typically have highly mixed and overlapping neural memberships, even when they are well-separated in the state space, Fig. 1a (middle column). In a special case of Hopfield networks, neurons are partitioned into largely disjoint groups with self-excitation within groups and inhibition between groups. In these winner-take-all networks, Fig. 1b, the attractor states consist of non-overlapping active groups of cells, Fig. 1a,b (middle columns).

**Continuous attractors.** How can one construct networks with a continuum of stationary attractor states? Weights that are invariant to reflection about their diagonal lead to the formation of discrete attractors, as we have seen. If the weights instead exhibit a *continuous* symmetry – for instance, if the weights are invariant to continuous shifts in neural locations – then the set of formed attractors will be related by the same symmetry and could thus form a continuous set.

The general principle for the formation of stationary continuous attractors is pattern formation [36, 37, 38, 39, 40, 41, 42]. Simple and spatially local competitive interactions lead to the emergence of rich stable spatial activity patterns – neurons with excitatory coupling between them become co-active, and suppress the rest of their neighbors through inhibition – a linear (Turning) instability [36].

The following three elements provide a solution to the inverse problem for forming stationary continuous attractors: 1) Nonlinear neurons with saturating responses or inhibition-dominated recurrent interactions with a uniform excitatory drive [7, 16, 43, 44] to keep network activity bounded. 2) Sufficiently strong recurrent weights with competitive dynamics in the form of local excitation or disinhibition with broader inhibition to drive spontaneous pattern formation via the Turing instability [36, 37, 38, 39, 40, 41, 42, 16]; these patterns become the attractor states. 3) Some continuous symmetry in the weights (a continuous weight symmetry is one where, as some variable is varied continuously, the weights remain invariant), such as translational or rotational invariance, Fig. 1c-d, to ensure a continuum of attractor states. These conditions are generally sufficient, but not strictly necessary, for the construction of continuous attractors (see Box on "Correspondences between attractor dynamics and anatomical layout"). If the continuous symmetry of weights is sufficiently corrupted, the continuous attractor will fragment into a discretized set of attractor states. Thus, the existence of stationary continuous attractors is fragile in the sense that it depends on the maintenance of continuous symmetries.

A special set of networks, which do not involve pattern formation to generate continuous attractor dynamics, are those with linear, planar, or hyperplanar attractors generated by neurons with linear or near-linear response functions. In circuits of linear neurons, the network feedback is a linear function of activity (**Wr** where **W** is the weight matrix and **r** are the neural activities), as is the activity decay (given by −**r**). Such networks can stablize non-zero activity states simply by tuning the strength of the feedback so that positive feedback cancels decay. The feedback matrix $W$ can direct feedback into the different dimensions of the state space; if feedback is directed largely along one dimension, the network can support a line attractor, Fig. 1e. If it is directed equally along two or more dimensions, it can support a plane or hyperplane attractor. To create long-lived attractors requires feedback to precisely cancel decay, thus the strength of network feedback must be finely tuned [9, 45], in contrast with pattern-forming continuous attractor systems.

**Non-stationary continuous attractors**: Large non-symmetric (and nonlinear) networks with strong connectivity generically exhibit limit cycle attractors or chaotic dynamics [46, 47]. Just as point attractors emerge generically in large networks with strong symmetric weights and bounded state spaces, chaotic attractors emerge generically in large recurrent networks with strong asymmetric weights. Adequate asymmetries are easily achieved if excitatory and inhibtory synapses emerge from distinct sets of neurons neurons [47], as biologically necessitated by Dale's law.

Despite the complexity of chaotic dynamics, these attractors are also highly structured in that they are typically much lower-dimensional than the number of neurons in the network [48]. Non-symmetric networks dominated by inhibition exhibit a single attractor at zero activity, though the flow towards the attractor in responose to perturbations can involve large transients in neural activation that temporarily move the state further away from the attractor [49, 50].

## The potential utility of attractors for computation in the brain

Networks with low-dimensional attractor dynamics exhibit myriad properties that can be vital for computation in the brain [4]. These include robust representation, memory, sequence generation, integration, and robust classification and decision making, ideas that have been extensively explored in the literature. In a later section, we will describe how, though attractor dynamics may be rigid and invariant as needed for the roles listed above, recent theoretical and experimental findings are beginning to reveal how these rigid constructions may also be exploited to perform flexible computation through reuse and re-combination across tasks.

### Representation and memory

A *representation* of a set of inputs means the assignment of inputs to representational states (a representation need not be injective), with the ability to reproducibly retrieve those states ('labels') when cued. Attractor networks provide a stable internal set of states that can be used for reproducible representation of discrete or analog variables, by mapping states in the world to the attractor states. One way to achieve this mapping is through a feedforward learning process that associates each external state with an internal attractor state, Fig. 2a.

An attractor network can exhibit two kinds of memory: The first is in the structure of the weights, which specify the set of all attractors. If these weights are specified through an input-driven learning process, this is a form of long-term memory about the inputs. The second is the ability to maintain *persistent activity* in a stationary attractor state: if a system with multiple stationary attractor states is initialized in one of them, it will tend to remain at or near the same state for some time. In other words, the activation levels of the neurons contributing to that state persist while the system remains in the state. This persistent activity response is thus a form of short-term memory of the input that initialized the circuit. If these persistent memory states can be activated without an explicit address, using just the content (or partial content) of the memory, they are content-addressable.

The short-term memory function of attractors depends on the prior formation of stable states through long-term plasticity: For instance, in Hopfield-like networks, states cannot persist if they were not first trained to be attractor states. Even models of short-term memory that are based on synaptic facilitation rather than persistent activity rely implicitly on prior long-term plasticity to construct recurrently stabilized neural ensembles that can be reinstated by random inputs [51]. In other words, these models cannot explain short-term memory for entirely novel inputs; however, combinations of attractors could enable more flexible short-term memory, as we discuss later.

---

[4] A system could theoretically be perfectly tuned such that every point in state space is a neutrally stable attractor, and thus the system has maximally high-dimensional attractor dynamics. However, because the robustness of attractor networks is related to the low-dimensionality of the attractor states as quantified in the subsections below, the system would lose most of its interesting computational properties: error correction/noise tolerance, nearest-neighbor computation, pattern completion and content-addressable memory. It could perform integration but with no robustness to noise.

**Denoising for fidelity of representation and memory**

If the representational states are attractors, then the representations are robust in the sense that they perform denoising: If the input cues or initial conditions reflect noisy or corrupted versions of an attractor state, the dynamics drive the state onto a point on the representational attractor, Fig. 2b (inset). When the attractors form a continuous manifold of dimension $K \ll N$, where $N$ is the number of neurons in the circuit, all noise in $N - K$ dimensions is erased. A noise ball of unit radius in $N$ dimensions (corresponding to random independent noise per neuron) has a projection of size only $\sim \sqrt{K/N} \ll 1$ along $K$ dimensions. If $K$ is low-dimensional, as is often the case, and $N$ ranges from $10^2 - 10^7$ as estimated before for common microcircuits, this constitutes a massive reduction in the sensitivity of the state to internal or input noise, Fig. 2b. Thus, most noise is rendered impotent.

Denoising due to attractor dynamics is especially important for memory maintenance, as otherwise noise-induced deviations would accumulate and grow over time. Discrete attractors continually erase all noise by mapping perturbed states back to the point attractor , resulting in zero drift. With continuous attractors as memory states, all noise orthogonal to the manifold is corrected, thus there is a net reduction of the effects of noise by the factor $\sim \sqrt{K/N} \ll 1$ [52, 53]. However, all states on the attractor manifold are neutrally stable so movements along the attractor are allowed. Thus, components of noise along the $K$ attractor dimensions are not internally corrected and cause an accumulating drift away from the inital state, with variance proportional to $KT/N$, where $T$ is the elapsed time [52, 16, 53, 54]. Thus, even continuous memory states can be well-stablized in sufficiently large attractor networks.

Although content-addressable long-term memory and error reduction can be instantiated through few-step feedforward computations [55, 56, 57] in place of attractor dynamics, recurrent attractor dynamics are indispensable for the generation of persistent activity states (and thus for short-term memory through persistent activity [58, 59]) and integration, as we discuss below.

**Robust classification**

When the attractors form a set with a discrete component (e.g. a set of point attractors or a set of continuous attractors), inputs that are not initially on one of the attractors will flow to one of the attractors and thus we may view the identity of the specific attractor to which the input flows as a classification of the input as a class represented by that attractor. The process can perform a pattern-completing nearest-neighbor computation, if the dynamics correctly drives non-attractor states to the nearest attractor states. In other words, the dynamical basins of attraction must align with the Voronoi regions of the attractor states, which is approximately the case for attractor networks operating well below capacity. This property deteriorates when attractor networks are pushed toward their capacity [60].

**Integration**

Single neurons integrate their inputs, but usually can only do this over the time-scales associated with their membrane capacitances, typically 10-100 milliseconds. Continuous attractor dynamics can allow neural circuits to integrate over much longer time-scales ( 1-100 seconds).

A non-linear continuous attractor network requires an additional mechanism to gain the functionality of an integrator: A way to shift the internal state along the attractor in response to an input that encodes changes in the external variable, Fig. 2d (left). Conceptually the simplest way to build a shift mechanism is by a *copy-and-offset* construction: construct multiple copies of the attractor network, each with slightly offset (asymmetric) weights in the sense that active neurons center their excitation or point of maximal disinhibition slightly offset from themselves on the neural sheet (e.g Fig. 1g (left) represents a slightly asymmetric version of Fig. 1c). The states in each such network will then form a limit cycle attractor, with patterns flowing in the direction of the asymmetry. If opposing copies are coupled together, the pattern is stabilized through a push-pull balance. A velocity input whose components project differentially to the copies will break the push-pull balance, allowing the more-active population of the moment to drive the pattern along its flow direction (cf Fig. 1g). Thus, the total direction and magnitude of the shift of the pattern, corresponding to movement along the attractor manifold, represents the time-integral of the velocity input to the network. This common principle unifies the mechanisms across diverse integrator models [61, 13, 14, 16, 62].

**Decision making**

If, instead of a velocity signal, the input to an integrator network consisted of temporally varying positive and negative evidence in support of two options [63], Fig. 2d (right) (or in the case of multiple options, evidence vectors instead of velocity vectors [64]), the network would integrate those inputs and thus perform evidence accumulation.

Decision-making can be viewed as a selection process applied to the integrator, based on a readout that detects when the integrator state has accumulated enough evidence and moved past a decision threshold [65, 54]. The selection process can be external to the integrator in the form of a readout circuit that detects such threshold crossings and outputs the decision; or it can be built into the dynamics of the integrator itself, in the form of a more-complex attractor landscape: the states evolve along a continuous attractor, but at some point the continuous attractor gives way to a pair of discrete attractors toward which the states flow, Fig. 2e. Neural winner-take-all (WTA) models implement such a hybrid analog-discrete computation [63, 17, 66, 67, 68, 64]. The parameters in WTA networks determine the balance between integration dynamics and competitive dynamics, and thus how well the network integrates later evidence (when the network is tuned to be a perfect integrator, its response to inputs is gradual and small amounts of evidence cause (reversible) flow along the continuous attractor manifold. In the case when competition dominates, the response to evidence is a fast flow toward one of the discrete attractors; beyond a point the flow is nearly irreversible, leading to rapid decision making and discounting of later evidence [69]).

Neural winner-take-all networks can accurately and rapidly (in $\sim \log(N)$ time) make the best decision among $N$ alternatives, even if the presented data are noisy (fluctuating over time around their means) [68, 64] and the number of options varies over orders of magnitude [64].

**Sequence generation**

Attractor dynamics can be important for stabilizing another long time-scale behavior: the generation of sequences. Robust sequences can be constructed as low-dimensional limit cycle attractors, in which high-dimensional perturbations are corrected, while along the attractor there is a systematic, periodic, or quasiperiodic flow of states [70, 71, 72, 73, 74]. The attractor property that affords ongoing de-noising is important for preventing spatial dispersion and temporal dissipation of the activity packet during sequence generation.

As for stationary attractor manifolds, the small components of noise along the limit cycle attractors are not correctable and lead to a gradual accumulation of drift, which for sequence generation is manifest as timing variability: the standard-deviation in the time of reaching the $T$th state in the sequence is predicted to grow as $\sqrt{T}$ for unbiased random drift along the attractor [53].

## Evidence of attractors in the brain

### Criteria for establishing attractor dynamics

The fundamental predictions of attractor models center on the state-space dynamics of the circuit, as first explicitly discussed and tested in a few papers [75, 9, 16, 76]: 1) That the system's states should be found localized at or around a much lower-dimensional set of states corresponding to the attractors in the state space. 2) That perturbations of the system should flow quickly back to the low-dimensional states. 3) That the set of attractor states – quantified by either direct characterization of the full state space or by the relationships between cells – should be invariant, persisting over time and after removal of tuned input, across conditions, across behavioral states, and even when there are induced variations in the mapping from internal states to external inputs [75, 16, 76]. 4) Integrator networks should further exhibit the property of isometry, in which lengths of coding space along a dimension are allocated to equal displacements along a dimension of the external variable. 5) Additional predictions of attractor dynamics models, that are not as fundamental in the sense that they are not theoretically necessary or sufficient but are nevertheless of high significance because they are highly supportive of the mechanisms of attractor dynamics, are anatomical and structural correlates: the existence of low-dimensional structures and symmetries in connectivity between cells.

Because attractor systems are characterized by their internally generated or autonomous dynamics[5], putative attractor networks are best tested in conditions that minimize external cues that are time-varying or tuned to provide localized inputs along the putative attractor.

Innovations in recording methods that made it possible to record multiple neurons simultaneously in animals performing naturalistic behaviors [77, 78, 79, 80], have enabled essential tests of these state-space predictions of attractor models. The newest methods provide activity data from $\sim$ 1000's of neurons within a circuit [81, 82, 83], making it possible to directly characterize the low-dimensional state-space dynamics of whole circuits [84, 85, 19, 86, 87].

When the attractor manifolds are < 3-dimensional, one can directly visualize them by projecting or embedding the high-dimensional state-spaces into ≤ 3 dimensions (using e.g. PCA, multi-dimensional scaling, tensor factorization, and other linear methods for projection, or Isomap, LLE, tSNE, VAEs, LFADS, nonlinear tensor factorization, and so on, for nonlinear embedding [88, 89, 90, 91]). These methods can also be useful when manifolds have dimension ≥ 3 but are topologically simple [86, 92]. For topologically non-trivial structures (e.g. rings, torii), especially those of dimension ≥ 3, topological data analysis methods become important[93, 94, 95, 19, 96, 97, 87].

---

[5]Attractor networks dynamics need not be used by the brain in an autonomous setting: inputs that drive attractor networks can be an important part of their function, for instance in integration and evidence accumulation. However, for the purposes of identifying mechanism, it is important that their dynamics are probed in an (effectively) autonomous setting

Testing predictions 1)-3) requires examination of the state-space structure of the population response, rather than the more conventional characterization of relationships (tuning curves) between cell activity and input or output variables. The most direct way to examine the state-space structure is to record enough cells simultaneously that it is possible to characterize the full state-space manifold [19, 96, 20]. However, the existence, stability, and invariance of lower-dimensional state-space structures (predictions 1)-3)) can be inferred from smaller samples of simultaneously recorded cells, by characterizing the invariant structure in pairwise cell-cell relationships, as has been successfully done in a number of studies [75, 98, 99, 76, 100, 101].

Predictions 1) and 2) are necessary but not sufficient for identification of recurrent attractor dynamics in a target network. First, if the behaviors and inputs are themselves low-dimensional, then any observed low dimensionality of the circuit states may be ascribed to the inputs and reveals little about intrinsic constraints imposed by the circuit. Second, even if inputs and behaviors are high-dimensional, a low-dimensional feedforward projection into the target network can generate low-dimensional target states and rapid erasure of high-dimensional perturbations. The *sina qua non* of attractor dynamics is prediction 3), which is that, because the states are internally generated and stabilized by strong recurrent connectivity, the population states and cell-cell relationships should be invariant when probed across time and across a wide and rich variety of input conditions including the removal of tuned input and across waking and sleep. In simple terms, the states observed in 1)-2) should be invariant across a broad range of conditions [16, 76].

Next, to the question of circuit localization: If a circuit exhibits the key signatures of attractor dynamics, does it originate these dynamics or is it a readout of some other region? Localization need not be a primary goal of establishing attractor dynamics: an important first step is to simply characterize whether the brain solves certain problems through attractor dynamics, regardless which circuits create these dynamics. Nevertheless, the persistence of activity states in attractors can lend a helping hand to localization efforts. If a region originates or is upstream of the attractor dynamics, but not downstream of it, then perturbations that succeed in altering its low-dimensional state should persist after the perturbing drive is removed [102].

As we illustrate next, theoretically-motivated analyses of population data have now firmly established that low-dimensional attractor dynamics are ubiquitous in the brain, across levels in the brain's hierarchy and across species.

## Discrete attractors

### Up and down states

The simplest example of nontrivial discrete attractor dynamics (i.e., beyond a single point attractor) is bistability. Bistable dynamics are a feature of cortical activity in the form of up and down states [103, 104, 105, 106, 4, 107, 108], in which the subthreshold membrane potential of neurons switches between a hyperpolarized state and a relatively depolarized one, with long persistence (100's of milliseconds to seconds) per state, Fig. 3a. The two states are relatively invariant over time, as seen in the relatively sharply peaked histograms (Fig. 3a; predictions 1), 3)), and despite presumed internal noise in the system the peaks are well separated, suggesting a relatively rapid corrective dynamics towards the two states (prediction 2)). There is little evidence of a critical contribution from cellular bistability in supporting these states, suggesting that it is a network-driven phenomenon involving self-excitation and global inhibition [109, 110, 111, 104, 105, 106, 4, 107]. Transitions are believed to be driven through adaptation (from up to down) and stochastic as well as external coordinating events (from down to up) [108]. Though these states and switches can occur in cortex without input from thalamus and striatum, they tend to be synchronous across cortex and striatum. Thus, the origin of up and down states may be highly distributed.

**Perceptual bistability**

Visual and auditory percepts including binocular rivalry, the Necker cube, and some auditory illusions [112, 113, 114, 115, 116, 117, 118] offer clear examples of bistability in neural processing. In these illusions, the brain (at the level of perceptual reports) selects one possible interpretation of an ambiguous input, often switching between possibilities. Though the phenomenon has long been known and studied, no localized circuit has been identified as the basis of perceptual bistability. Indeed, some percepts may involve top-down activation and modulation of activity across many brain areas [116], suggesting once again a widely distributed circuit for bistability.

**Bistability in a premotor area**

Recent studies identify and localize discrete attractor dynamics in a mouse premotor area, the anterior lateral motor cortex (ALM) [119, 120, 121, 122]. In a cued 2-alternative delayed response task ALM neurons exhibit persistent activity over a $\sim$ 1s delay period. During the post-cue delay period, activity evolves toward one of two states that guide the response, Fig. 3b (prediction 1)). The delay-period terminal states are similar for cues from different sensory modalities [123] (partial test of prediction 3)). ALM perturbations during the delay are either erased (corrected) by the circuit (Fig. 3b, top) or drive a jump to the opposite state (Fig. 3b, bottom), which results in the animal making the wrong action, suggesting a bistable switching dynamics similar to the mechanisms in either Fig.1b or Fig.2e (prediction 2)).

Given the long training time required for the task and the resulting tailoring of the ALM dynamics to the specific task structure – bistability for a two-choice task – it is likely that this system acquires its dynamics through slow plasticity and thus that the network's recurrent structure is malleable in adult animals. New results showing the existence of small ($\sim 100 \mu$m scale) locally recurrent clusters of neurons ALM that can maintain persistent responses to microstimulation [124] may provide experimental evidence of the theoretically posited mixed modular networks (below) hypothesized to support robust and high-capacity memory states [60].

**Discrete multistability**

Hopfield networks and winner-take-all (WTA) networks are models of multistability beyond bistability[6]

To date, there are somewhat less direct data and exhaustive analyses to establish discrete multistability as a circuit-level brain process, in comparison to the evidence for continuous attractor networks (described next). However, there are many likely candidates systems and brain regions with dynamics suggestive of and consistent with discrete multistability, at least of the special case of WTA attractor dynamics, including in mammalian hippocampus and auditory cortex, and the fly and mamalian olfactory system [133, 134, 135, 136, 137, 138]. In particular, many of these circuits exhibit global inhibition that clearly narrows and refines activity in the circuit (prediction 5)), Fig. 3c top versus middle, and also show evidence of selective recurrent excitation that leads to multiple distinct and stably correlated input responses in distinct subpopulations of cells, Fig. 3c, middle versus bottom [133, 134, 135, 136, 137, 138]. In our view, it is likely that these circuits exhibit multiple discrete attractor states but quantitative testing of predictions 1)-3) and direct demonstration of these states as stable and invariant remains an important future direction for characterizing these circuits.

---

[6]WTA networks [125, 126, 127, 67, 128, 129, 130, 131, 132] may be viewed as a special case of Hopfield networks, and bistable switch networks are a special case of WTA networks. As noted earlier, both can express mulitple discrete attractor states, but while the Hopfield network attractors have highly mixed and overlapping neural membership, WTA attractors consist of activity in largely disjoint groups of neurons.

## Continuous attractors

### The oculomotor integrator

The oculomotor integrator, together with the HD circuit, was one of the first systems in neuroscience to be studied theoretically [8, 139, 9] and experimentally [140] as a continuous attractor network – specifically as a line attractor, Fig. 1e. This network, presynaptic to the motor neurons that control horizontal eye position, is highly conserved across vertebrates, from fish [141, 140] to primates [142, 143]. It integrates pulse-like saccadic eye movement command signals to generate step-like stable muscle tension command signals (Fig.4a) that persist autonomously at various graded activity levels after removal of the movement cue and even in the dark in the absence of visual feedback (Fig.4b; prediction 3)) and thus enable stable gaze fixation at various eccentricities. Saccadic inputs knock the system slightly off the linear response states, but the neural responses rapidly decay back towards the persistent firing states (prediction 2)). Remarkably, the same system also integrates smooth head velocity signals to permit gaze stabilization during head movement. Integration functionality is a network-level rather than single-cell process: single neurons do not generate persistent responses to transient current injections, Fig. 4c (inset), while decreasing network feedback through synaptic blockers reduces the time-constant of integration and results in a leaky integrator [144], Fig.4c. It is possible to induce a reduction or increase in network feedback through training with a virtual surround that generates an artificial retinal slip percept, Fig.4d, showing that the system is capable of error-driven fine-tuning to maintain a high degree of persistence [145]. Finally, a recent EM reconstruction [146, 147] finds recurrent synaptic interconnectivty between integrator neurons, with excitatory connections between ipsilateral neurons and primarily inhibitory contralateral projections, in excellent agreement with line attractor models of the oculomotor circuit [9], Fig.1e (prediction 5)).

### Head direction cells

Some of the earliest experiments to suggest the existence of low-dimensional continuous attractor dynamics were done in the rodent head-direction (HD) circuit[98, 75, 148], Fig.5a,b. The HD circuit in mammals maintains an updated internal compass estimate of heading direction (relative to some arbitrary external reference) as animals move around. It does so by integrating internal rotational velocity estimates during navigation and incorporating information from external cues [149, 150, 151, 152, 153]. The HD circuit is modeled by the ring attractor network [10, 61, 11, 154, 13, 14], Fig.1c, g (left). Before large population recordings became available, cell-cell correlations established that the network states remained invariant on a very low-dimensional manifold across environments [98, 75, 148], Fig.5a (predictions 1), 3)). Recently, the complete set of states of the several thousand neuron-sized mammalian HD network was shown to consist solely of a 1-dimensional ring, Fig.5b [19, 96] (prediction 1)), revealing that the brain has completely factorized its navigational representations to dedicate a circuit only to head direction. Further, intervals in the state-space ring manifold map isometrically to intervals of head direction (prediction 4)), as evidenced by a close match between the isometrically parameterized internal ring states and the measured head direction, Fig. 5b (inset, right).

Natural perturbations away from the ring flowed back to it, Fig.5d [19] (prediction 2)), and the ring manifold was invariant across waking and REM sleep, Fig.5e [19, 96] (prediction 3)). These findings explicitly validate the most fundamental predictions (predictions 1)-3)) of ring attractor models and continuous attractor-based integrators (predictions 1)-3) with 5)), providing arguably the most direct and compelling evidence of continuous attractor dynamics in the brain.

In a striking example of convergent evolution [155, 151], *Drosophila* compute HD estimates using apparently very similar dynamics [156, 157, 153, 152]. The fly neural compass circuit is topographically organized such that the neuropil forms a physical ring-shaped structure in the ellipsoid body, with a local moving activity peak that tracks head direction as the fly turns, Fig. 5f. Another notable advantage of the fly circuit in the effort to characterize its mechanisms is that the number of neurons is small and their morphology and connectivity has been fully traced [158], Fig.5g. This detailed view of the circuit permits quantitative, not just qualitative, comparisons with ring attractor models. The combined activity and connectivity data reveal that the fly HD system implements the copy-and-offset double-ring network architecture proposed for velocity integration [159, 14]. The actual dimensionality of the fly HD circuit and its full state-space dynamics remain to be characterized; even though the circuit is organized physically as a ring network, recent evidence suggests that the insect HD circuit may be involved in performing 2-dimensional path integration as well [160, 161], and thus unlike the ADn network in mammals, may not be confined to a 1D ring of attractor states that fully factorizes head direction in its representation of spatial variables.

Finally, the HD system can be re-anchored and reset based on tuned external cues [152, 153], which can change the orientation tuning curves of cells and moment-by-moment firing rates of cells in a way that remains consistent with prediction 3) for attractor dynamics.

**Grid cells**

Grid cells encode spatial location though a regular triangular-lattice discharge pattern that tiles explored 2D spaces [162], representing 2D position as a set of spatially periodic 2D phases [163, 129]. They update their states while moving in the light and the dark [162], presumably based on motion cues. Continuous attractor models [15, 164, 16, 165] predict that the population states of a module – a set of grid cells with a common period – should be confined to merely 2 dimensions regardless of environment and behavioral state [16], forming a manifold that is topologically a torus, Fig.1d (rightmost column).

Indeed, grid cells from one module ("co-modular cells") have identical periods and orientations and all possible 2D phases, suggesting a 2D set of states [166, 76] (prediction 1)). The relative firing phase and relative grid parameters of pairs of co-modular cells is tightly conserved even as the spatial tuning and spatial phase of single cells varies across time and across familiar and novel environments (Fig. 6a) [166, 76], across the dimension of the spatial environment (6b) [167], and despite environmental rescaling that leads to large deformations in the spatial tuning of grid cells [76] (prediction 3)). Moreover, the detailed cell-cell relationships (whether a pair of cells is co-active, active in quadrature, or active fully out of phase) that are seen in waking exploration are conserved across overnight REM and non-REM sleep for grid cells but not place cells, Fig. 6c [100, 101], establishing that the low-dimensional states are autonomously generated (prediction 3)). These findings established that the structure of the grid cell response is very low-dimensional on a population level, invariant across environments, time, and behavioral states, and internally stabilized and autonomously generated – validating the fundamental predictions [16] made by continuous attractor grid models. Most recently, these findings were reproduced by a direct visualization of the state-space manifold, Fig6e made possible by large population recordings of grid cells across waking and sleep that confirmed the toroidal state-space structure of co-modular cells [20].

A corollary of these findings is that the grid cell response is *not* derived from upstream place cell inputs since place cells remap across environments and during sleep while grid cells retain their population structure (Fig. 6c); this corollary belies and is inconsistetnt with models in which the place cell response is primary to grid cells [168, 169, 170], as shown in [100].

Given the preserved internal structure and autonomous dynamics of grid cells across states, time, and environments, it follows that various deviations in the spatial tuning curves of grid cells from equilateral grid-like responses in 2- and 3-dimensional spaces [171, 172, 173, 174, 175, 176] likely result from variability in how the invariant internal states are driven by and mapped to external cues and states: for instance through altered velocity estimation [16, 76] or feedforward inputs that shift the phase of the grid cell network [177, 178, 179]. This has been verified in the case of the expansion of grid cells in novel environments [76] and is almost certain to hold – given the preponderance of evidence of internal grid stability [166, 76, 180, 100, 101, 20] – when tested in various other conditions that report grid deformations as well [172, 173, 174, 175, 176, 181].

In sum, the HD and grid cell systems confirm that the same pattern formation principle – based on local excitation or disinhibition, with broader inhibition – that is pivotal for morphogenesis in plants and animals [39] is also fundamental to the genesis of stationary continuous attractor states for computation and representation in the brain.

**Graded working memory networks**

In monkeys trained to make saccades to previously cued targets (selected from a set arranged in a circle), neurons in PFC and PPC exhibit persistent activity tuned to the direction of the initial cue, across the delay period after cue removal (predictions 1), 3)) [182, 183]. Analysis of delay-period PFC activity [85] in a population of simultaneously recorded shows that the delay period activity bump moves apparently randomly along a 1-dimensional manifold, with the characteristics of a diffusion process, so that the variance in the location of the bump grows linearly with time, as predicted by continuous attractor models [52, 16, 19], but the bump profile remains largely invariant over the duration of the delay (predictions 1) and 2), assuming that the diffusive process is indicative of natural noise-driven perturbations of the system). The bump movement predicts subsequent behavioral errors [85], suggesting that these states are repositories or readouts of the memory.

The need for extensive training on the task and the observed tailoring of the attractor states to this specific multi-cue task suggests that this attractor formed through learning in a flexible system rather by (re)using a genetically pre-specified circuit. Given the apparent malleability of this attractor network, we might therefore also expect a loss of the neural correlation structure if the animal is subsequently trained on other tasks, unlike with the grid and HD cell networks.

**Attracting limit cycles and trajectories**

The central and peripheral nervous systems contain numerous instances of periodic dynamics, ranging from the spiking of single neurons [184, 185] to circadian rhythms and sleep cycle generation [186], to rhythmic activity in motor circuits. While linear oscillators have amplitudes set by the initial condition, attractive limit cycle oscillators have an intrinsic and invariant amplitude. Thus, not all systems with oscillatory behavior are limit cycle attractors: oscillations that decay or grow over time or whose long-term amplitude or frequency changes after a transient perturbation are not limit cycles. Driven (non-autonomous) systems may exhibit limit cycles because of their inputs rather than intrinsic attractor dynamics [187].

Many of the oscillations noted above maintain a fixed amplitude, and because of their strong functional imperatives for robustness to perturbation are almost certainly generated through attractor dynamics. Particularly well-characterized examples are central pattern generators (CPGs) in motor circuits of the peripheral nervous system, that drive swimming, crawling, walking, breathing, and digestion, and differ in specifics across species but have common principles of mechanism and operation, including high robustness [188, 189]. CPG circuits typically integrate external feedback but have been shown to be able to operate in isolation without external drive [190].

Given the sizeable literature on these topics, we refer the reader to some excellent papers and reviews [191, 192, 193, 194, 195, 186].

## Departures from low-dimensional continuous attractor dynamics

Not all circuits hypothesized to exhibit low-dimensional attractor dynamics appear under further experimentation to do so, or currently lack sufficient evidence to establish such dynamics within the circuit. We discuss three potential examples.

### Orientation tuning in V1

The circuit of simple cells in V1 satisfies some key properties of ring attractor networks [11]: V1 and V2 cells exhibit strong orientation-tuned responses to real and illusory edges in the visual world [196, 197, 198], and population-wide V1 spontaneous activity during sleep is correlated with these tuned population coding states [199]. While these suggests that illusory edge responses may be driven by self-generated attractor dynamics, they tend to occur after a longer latency than real edge responses, making them more likely to be driven by top-down inputs rather than within-V1 dynamics. Next, moving an attractor state along a continuous attractor manifold requires strong inputs and is slow [200, 201]. These features seem inconsistent with the imperatives of a perceptual system to respond rapidly to changes in input [202], and the dynamics of state fluctuations in sleep appear relatively rapid compared to attractor time-scales [199]. These observations seem to lend more weight to models in which the circuit response is dominated by feedforward drive [196, 203], possibly with recurrently generated but fast non-normal amplification processes [49, 204]. More quantitative characterizations of response speed will be important to draw clear conclusions about competing models for V1 circuit dynamics.

### Place cells

Place cells form stable and detailed representations of familiar 1-2 dimensional spatial environments [205], which can persist in the dark [206] and for short intervals after the animal has fallen asleep [207, 208]. In any particular 2-dimensional environment, the population response lies on a low-dimensional manifold in state-space [86]. Accordingly, the place cell circuit has been modeled as a 2-dimensional continuous attractor network [209, 210] or as a superposition (with overlapping neural membership) of a discrete number of such 2-dimensional continuous attractors, each representing a different environment[211]. However, the capacity limitations of generating multiple high-resolution maps within one homogeneous attractor network are severe [212, 213, 163, 214]. And, unlike grid cells, place cells do not, across long sleep bouts, exhibit the spatial correlations measured in awake exploration Fig. 6c [100, 101, 207]. Even during waking, cell-cell correlations are not preserved across environments because of the phenomenon of remapping [215, 216]. Like V1 neurons, place cells might be better described as deriving their tuning by forming conjunctions of multiple feedforward inputs, including from grid cells and cells that encode external cues like borders, landmarks, and reward sites [217, 218, 129, 214]. At the same time, place cells exhibit sequential activation of previous trajectories during activity replay [208, 219, 220, 221], which is hypothesized to be generated by recurrent connections in CA3, suggesting that recurrent and feedforward dynamics collaborate in the generation of place cell states; recent models are beginning to capture this interplay [210]. Closing the book on the question of autonomous low-dimensional dynamics in the far more complex response of place than grid cells requires more detailed experimentation, analysis, and modeling.

**Motor cortical trajectories**

Finally, recordings of motor cortical activity during stereotyped primate arm movements reveal the existence of stable low-dimensional trajectories [84, 222, 223, 224, 225], similar to the trajectories in state space originally characterized in olfactory circuit responses to odors [226]. Limit cycles and other low-dimensional attractors have been hypothesized to play a key role in cortical movement generation [227, 228]. The behaviors typically performed during neural recording are themselves restricted to be stereotyped and low-dimensional, thus it remains unclear whether activity would remain equally low-dimensional across a richer set of behaviors (e.g. over the set of all possible arm movements). Recent evidence from perturbation experiments [187] suggests that neural trajectories in motor cortex during skilled movements are driven by input from the thalamus, and thus that the circuits for motor pattern generation in the central nervous system might be distributed across multiple brain regions. Characterizing the intrinsic dimensionality of motor cortical activity, and determining whether the command to make more-complex motions involves multiple upstream or distributed primitive attractors, remain important open questions for both clinical brain-machine interfaces and neuroscience.

# Flexibility despite rigidity: modern glimpses into the broader potential of attractor networks

Above, the key predictions and experimental validations of attractors in the brain hinged on their invariance, or rigidity, across time and conditions. The identified attractor states were highly structured and low-dimensional. The weight symmetries and asymmetries underlying these states were precisely tailored to the specific tasks performed by the systems. These properties appear to run counter to a key desideratum for representation, memory, and computation in the brain: flexibility.

Recent experimental and theoretical work are beginning to shed light on how the brain might solve the perennial conundrum of stability versus flexibility through attractor networks: the low-dimensional and rigid attractor states might be reused and recombined to create versatile and efficient systems for novel situations.

**Exploiting integration for rapid representation: reuse of continuous attractors**

Strikingly, all established stationary continuous attractor networks in the brain are also integrators. This seems surprising given that not all continuous variables represented in the brain need be accumulation of evidence or navigation-like variables. Here we discuss how, even for just the problem of representation, the functionality of integration could serve a vital role by enabling rapid construction of new representations.

Building a representation (mapping values of an external variable to the internal states on a continuous attractor) as in Fig. 2a can proceed by painstaking construction of a large set of associative feedforward correspondences: Visit each external state and associate it with an attractor state. Building this lookup table requires full exploration of the space. By contrast, if the attractor is an integrator, only two feedforward correspondences must be built: identify one value of the external variable with one internal state – an anchoring process – then associate the velocity signal with the shift mechanism in the integrator through a learned feedforward projection that is independent of location on the attractor or in the external space, Fig. 2f [229]. The circuit can now automatically generate appropriate and consistent representational states for future and previously unvisited values of the variable based on displacements. This mapping is rapid and does not require exhaustive exploration of the space or reconfiguration of the recurrent attractor circuit, a form of generalizable and rapid learning [230, 231, 229] and *zero-shot memory state construction* [229]. Moreover, an integrator network can correctly infer the current state upon returning to it along a previously untraversed path (novel trajectory), a form of *zero-shot inference* [231, 229, 232].

This rapid integration-based mapping process permits another use: A single attractor can be easily *reused* to represent mutiple variables, Fig. 2f: By simply adding another anchor point for another variable $Z$ and driving the shift mechanism with velocities related to changes in $Z$, the network can switch from representing variable $X$ to de-novo representing $Z$, or can alternate between representing $X$ and $Z$ without any reconfiguration of the attractor itself [229]. Consistent with this idea, it appears that the brain (re)uses grid and place cells when navigating through both the spatial environment and through non-spatial cognitive domains [233, 234, 235].

**Multiple modular attractors for high-capacity representation**

In general, a fully connected symmetric attractor network of $N$ neurons permits the construction of $\sim N$ chosen attractor states (where the notation $\sim$ refers to the functional scaling with $N$ with potential prefactors that do not depend on $N$), a result established by a large body of work from statistical physics [236, 237, 238] and information theory [239]. The result is independent of learning rules [239, 236] and not qualitatively altered by adding hidden neurons [60]. It also applies to (near-)continuous attractors, for which the total number of distinguishable states (resolution) on an attractor manifold also scales in the same way.

A $D$-dimensional attractor with resolution $\sim P$ per dimension would thus require a number of neurons that grows exponentially with dimension, $N \sim P^D$ (this is one aspect of the *curse of dimensionality*). For 2-dimensional space at a resolution of 10 cm per dimension, the full population of $\sim 10^6$ rodent hippocampal cells – which have been hypothesized to function as a Hopfield-like associative memory – could at most represent a combined 100m$^2$ area [163]. Similarly, the number of cells in hippocampus is vastly smaller than the combinatorially large number of sparse coding states of cortical neurons in rodents and humans, which might set the scale for the number of items to be stored in memory. These arguments point to the need for much higher-capacity representations and memory than possible with fully-connected Hopfield-like networks.

Modular attractor networks permit efficient construction of a large number of representational and memory states in neural networks [163, 129, 240, 241, 242, 243, 60, 214]. If $M$ modular subnetworks have $\sim N$ discrete attractor states each, and these can update independently (uncoupled modules), the combined system expresses a set of $\sim N^M$ states, exponential in the number of modules. Though these states are not attractors, it is possible to couple together these subnetworks to generate exponentially many attractor states so they each have reasonable-sized basins and are thus robust [244, 241, 242, 60, 57], Fig. 2. (By contrast, randomly connected Hopfield networks also typically have exponentially many fixed points, but the basins are often small.) Similar ideas have been applied in the temporal domain to show how networks might support exponentially long activity sequences [245].

If each module expresses a continous attractor of dimension $K$ and the subnetworks are independent, their combined states define an $MK-$dimensional manifold (with no error correction between states on the manifold), solving the curse of dimensionality for representing higher-dimensional variables while maintaining a structured representation for the variable that goes well beyond random combinatorial codes [163, 129, 246]. These subnetworks can be coupled together through their shift mechanisms, forcing a certain fixed relationship between modules in how their relative states update [243], in which case the coupled system exhibits one $K$-dimensional attractor but with large capacity, containing $\sim e^M$ states per dimension.

In short, modular subnetworks can work together to greatly expand representational capacity in terms of number of attractor states while also maintaining a large denoising capability [244, 241, 242, 60, 57, 163, 129, 246, 243]. However, the vast set of attractor states made possible by these coupled-module constructions are a rather structured set pre-defined as combinations of the states in the individual modules, rather than being arbitrarily specified as the states are in a standard Hopfield network. They do not directly encode user-defined patterns as the attractor states, and thus do not violate the capacity limitations of neural networks [236, 237, 238, 239]. A critical question is how high-capacity and robust but structured sets of attractor states could be leveraged for general memory and computation. Three recent works have begun to address this question, showing that structured attractor states can be leveraged for robust labeling and action selection [247], robust classification [248], and as a component of a heterogeneously structured general associative memory that exhibits smooth degradation instead of catastrophic memory loss as more patterns beyond capacity are added to the network sharma2022content.

**Mixed modular codes for flexible representation**

Finally, $M$ modular subnetworks that are each integrators in $K$ dimensions can be (re)used without any rewiring of recurrent weights to represent and store inputs of any input dimension $\leq MK$, Fig. 2h, using a *mixed-modular coding scheme* [229]. This scheme combines five concepts: Rapid representation learning with the integration mechanism, the reuse of the same attractors for different variables, the capacity of modular attractors, the fact that a high-dimensional variable can be represented unambiguously by multiple independent lower-dimensional projections, and the fact that random projections tend to be independent.

In mixed modular coding, movement along each dimension in the external space is randomly projected to the shift mechanisms of all modular integrators. Every module is thus involved in representing every input dimension – a form of holographic representation [249, 250]. If the number of input dimensions $D$ is smaller than $MK$, all input dimensions are represented without information loss, and excess module capacity ($MK - D$ excess dimensions) is automatically convertible into extending the coding range or resolution for each of the *D* input dimensions, instantly trading off the number of represented dimensions and the dynamic range of each represented dimension without recurrent plasticity.

## Looking ahead

The theory of attractor dynamics in the brain has provided a powerful and unifying conceptual framework for understanding integration, representation, memory, error-correction, and efficient learning and inference in the brain. The experimental effort to study candidate attractor circuits and test their predictions has been a fertile field of research, and population-wide physiology techniques have led to breathtaking direct visualizations of attractor dynamics at work in the brain.

The theory is also proving to be a powerful tool in interpreting how artifical neural networks (ANNs) solve complex tasks. ANNs trained to robustly solve memory, integration, and decision-making tasks in domains as diverse as spatial navigation, vision, and language, develop attractor dynamics [251, 252, 253, 254, 255], suggesting that not only are attractor networks able to solve such problems but might be necessary when the computing elements are memoryless neurons. Further, equipping networks with preconfigured attractor networks can help to produce faster, more data-efficient and generalizable learning [230, 229]. Because ANNs can be trained on complex tasks and then fully examined after learning, they will potentially more readily contribute to the next chapter in our understanding of how continuous attractor networks can interact and combine with other mechanisms to allow the brain to solve richer problems associated with intelligence.

Notable mechanistic questions about attractor networks also remain open, including: Moving away from the high-firing-rate asynchronous spiking regime [256, 257] to better understand whether low-firing-rate synchronous spiking networks might support attractor dynamics – and thus permit a combination of fast time-scale dynamics like spike synchronization and oscillatory phase dynamics [256, 258, 259]. For continuous attractors, understanding how the brain deals with the problem of fine-tuning in linear networks or the imposition and maintenance of a continuous symmetry across neurons remains unknown and ripe for resolution [34, 260].

A few continuous attractor development models show how they could emerge simply through unsupervised associative plasticity [10, 211, 178]; others are based on combining feedback of known or plausible error signals with neural activity in relatively simple learning rules [261, 10, 262]; the rest train networks on a high-level goal with error backpropagation, combined with other constraints on architecture or the form the solutions should take [263, 264, 252, 253, 230, 265, 254]. These models are incomplete for different reasons: the unsupervised models require uniform exploration of the input variable space and suppression of recurrent weights during their training; the backpropagation models do not offer an account of how the loss functions, learning, and additional constraints might be generated in biological systems.

There is much left to do in the field and an exciting vista ahead. On the experimental side, tools for high-resolution population-level neural recrdings and perturbation across multiple brain areas [266, 82, 83] let us peer further and deeper than ever. On the theory side, future developments will help us conceptualize how such circuits could help underwrite intelligent computation through the formation, interaction, and reuse of multiple low-dimensional structures.

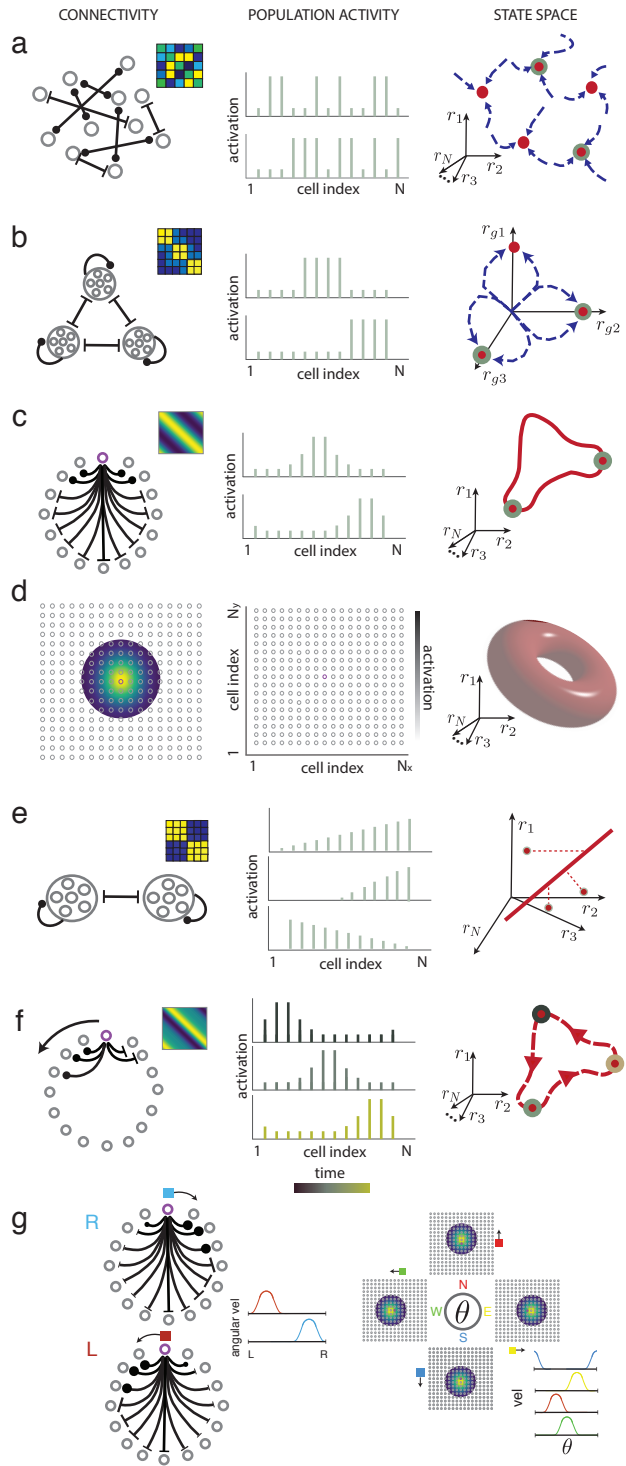

> **Correspondences between attractor dynamics and anatomical layout, and weight symmetries in models and biology**
>
> Anatomical topography, in which functionally similar neurons are near one another, is neither a necessary nor sufficient condition for the existence of an attractor: Any low-dimensional attractor network is mathematically unchanged if all weights are preserved but neuron locations are scrambled. However, if the network is merely a spatially scrambled version of the idealized model, then the symmetries of the weight matrix can be revealed after an appropriate reordering of the neurons. An advantage of anatomical topography from a biological perspective is that it can reduce the complexity of development, in that wiring decisions can be guided by spatial proximity rather than depending entirely on activity or other target cell signalling mechanisms; for instance, the locally competitive interactions of grid and HD circuit models could be largely constructed through local arborization. It also reduces overall wiring length in the mature circuit [275]. However, a circuit with ≥ 3-dimensional dynamics that are represented in an unfactorizable form cannot be embedded topographically in a 2-dimensional cell layout, which limits the utility of topographic layouts for circuits representing higher-dimensional manifolds.
>
> Second, the posited weight symmetries in simple models of attractors need not exist in a biological instance of the circuit with the same dynamics: If low-dimensional attractor dynamics is only needed downstream of the size-$N$ recurrent network that generates the dynamics, in a set of $M < N$ neurons, then the symmetries required for continuous attractor dynamics can be spread across the recurrent and readout weights and the recurrent weights alone without taking into account the readout weights will not reflect the relevant symmetries [253]. These considerations suggest a hypothesis for circuits with ≤ 2-D continuous attractors: Evolutionarily conserved circuits that do not require extensive early experience [276, 277] should be topographically organized. We might thus predict that the circuit that originates HD signals in mammals should be topographically organized. By contrast, if the low-dimensional dynamics only emerges on the basis of activity-dependent plasticity with repetitive training, we may not expect the circuit to be topographically organized (or even localized to single brain regions).
>
> Remarkably, despite these caveats and in a beautiful example of the predictive power of simple theories in neuroscience, the recent empirical evidence from the anatomy of the zebrafish oculomotor integrator and the fly HD circuit show that nature has used precisely the hypothesized constructions proposed in simple circuit models to build some integrator networks.

Figure 1 *(preceding page)*: **Mechanisms of attractor formation.** In all plots, open gray neurons represent neurons, connections between them are excitatory (black lines ending in bars) or inhibitory (black lines ending in circles). Left column: layout of neurons and connections; connectivity matries shown as inset, with blue to yellow colors indicating strongly inhibitory to excitatory interactions. Middle column: examples of stable population activity patterns. Right column: state-space views of population states and dynamics. Red circles with gray-green ring indicate the activity states shown in middle column. (a) A network with dense symmetric connections determined by associative Hebbian learning on a set of input patterns (middle) stores them as stable attractor states. This defines a Hopfield network. (b) Disjoint groups of neurons interacting through within-group excitation and across-group mutual inhibition leads to group winner-take-all dynamics. Stable states are any patterns with only one winning group (the state-space plot collapses all activities of neurons in group $gi$ along the axis $r_{gi}$). (c) Neurons arranged in a ring with global inhibition and either local excitation or a lack of local inhibition combined with uniform excitatory input to all neurons produces localized activity bumps (middle) as the stable states. Bumps may be centered anywhere on the neural ring, defining a near-continuum of attractor states that form a ring in state-space (right). (d) Neurons arranged on a two-dimensional sheet, interacting through local inhibition and either center-excitation or a lack of inhibition near the center together with uniform excitatory input to all neurons results in a pattern of multiple, periodically spaced activity bumps (middle). Any two-dimensional phase shift of the periodic pattern upto the lattice periodicity are distinct but equivalent stable states, then the states repeat; thus these are predicted to form a torus in the state-space. (e) Two neuron groups with in-group excitation and across-group inhibition, precisely tuned interaction strengths, and quasi-linear neural fI responses can counteract activity decay in the network and produce persistent activity over a continuum of activity levels in the two populations, defining ramp-like population activity states and a line of attractor states. (f) Neurons arranged on a ring with asymmetric connections that bias neural activity to flow in a particular direction (middle) The network forms localized activity bumps that sequentially move around the ring in that direction (right) The state space contains a limit cycle. (g) The copy-and-offset mechanism for constructing integrators, illustrated for the ring (left) and grid (right) attractor circuits. Each network copy receives velocity inputs tuned to the corresponding shift direction.

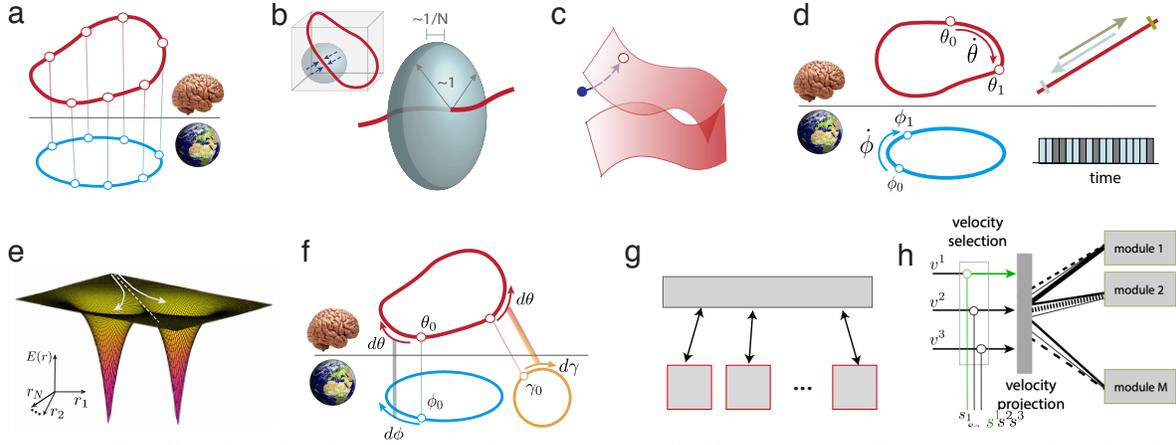

Figure 2: **Utility of low-dimensional attractor networks.** (a) Persistent and stable states generated by attractor networks (red) can be used to represent and remember external variables (blue) by constructing an appropriate mapping between them (vertical lines). (b) Noise-robustness: attractor networks error-correct by mapping noisy states to the nearest attractor state [267]. Main: $N$-dimensional noise drawn from the unit sphere centered on a 1D attractor has a projection strength of only $1/N$ along the attractor: in this counter-intuitive high-dimensional geometry, a ball is more like a pancake with the attractor orthogonal to the large dimensions [19]. (c) Flow to the nearest (continuous or discrete) attractor can perform a nearest-neighbor computation and thus perform classification: e.g. the two attractors may represent "cat" and "dog" perceptual manifolds, and the blue dot a specific input data point. (d) Left: Continuous attractors can become integrators if velocities or movements in the external space are inputs to the network and induce proportional shifts in the internal attractor state: The current state on the attractor is then the integral of past velocity inputs relative to the starting state. Right: if the input to an integrating attractor consists of temporally varying evidence pulses (bottom, evidence about one option in blue and evidence about the opposing option in khaki), these will move the state on the attractor (top) so its current state reflects the integral of the total evidence.(e) The energy landscape of a combined integration and decision making network: inputs push the state left or right, and as it integrates, the network state also moves toward one of two discrete attractors (left and right; white arrows: two sample trajectories). Arrival to the neighborhood of one of the discrete attractors is a decision point [63, 64]. (f) An integrator can be quickly re-purposed to represent multiple different and new external variables simply by yoking its velocity shift mechanism to different external velocities cues by feedforward learning. It also does zero-shot learning and inference: Given an initial state and an input velocity trajectory, it will generate a self-consistent representation for the current state even if the trajectory if different and new each time [229, 231, 232]. (g) A set of (continuous or discrete) attractor subnetworks (red boxes at bottom) can interact bidirectionally with a shared network to form a high-capacity attractor network [241, 242, 268, 60]. (h) Mixed modular representations can enable representation of inputs of different dimensions by resuing the same attractors of fixed dimension each. Velocities ($v_i$) from external spaces of potentially different dimension are selected by a set of selection signals ($s_i$). The selected velocity (green) is routed through random projections to a set of $M$ modular integrator networks of dimension $K$ each. This kind of mixed modular circuit can interchangeably represent a variety of input spaces of dimension $D \leq MK$ while smoothly trading off resolution for dimension [229].

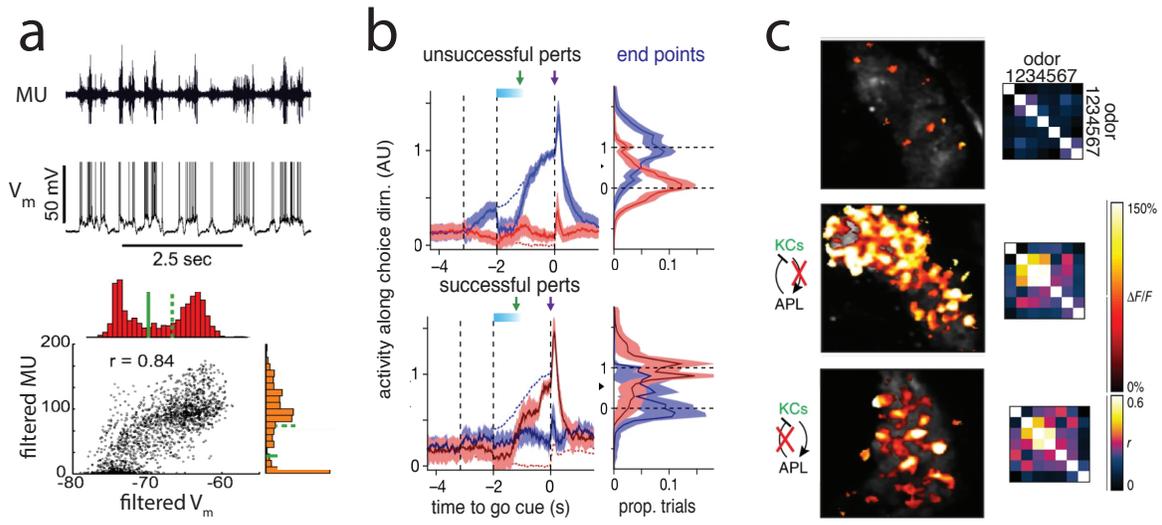

Figure 3: **Evidence of discrete attractor dynamics in the brain.** (a) Multi-unit (above) and single-unit (middle) activity during cortical up- and down-states show signatures of bistability (clusters and histograms at bottom). Reproduced with permission from [269]. (b) Delay-period dynamics in rodent premotor area ALM during a binary decision task before the animal can make a motor report of its decision appears to converge to one of two discrete end points (blue and red curves and histograms, top). Perturbations are either robustly ignored (top), or flip the dynamics so that the end points are reversed (bottom). Reproduced with permission from [123, 119]. (c) Evidence of all-to-all inhibition and competitive winner-take-all recurrent dynamics in the fly olfactory system: Kenyon cell (KC) responses to odors, with input from the globally projecting APL inhibitory neuron, are sparse (top left, Ca fluorescence response to odor IA) and decorrelated across odors (top right); blockage of KC drive to APL or APL inhibition to KCs results in dense and correlated odor responses (middle, bottom). Reproduced with permission from [137].

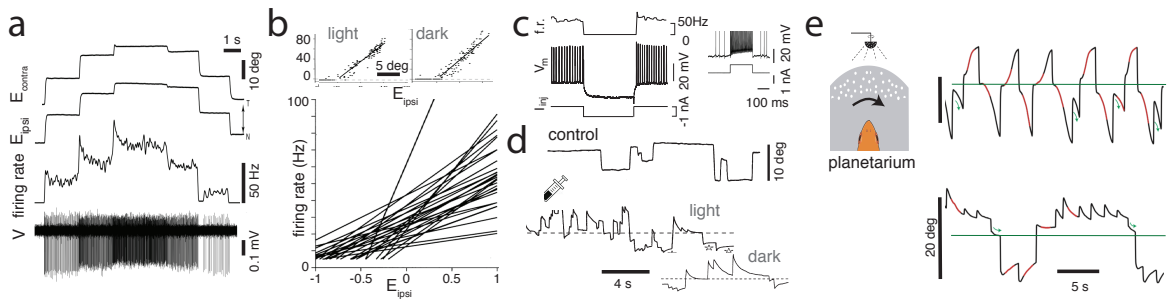

Figure 4: **Linear attractor dynamics generated by newtork feedback in the oculomotor integrator.** (a) Stable horizontal gaze during fixation at different angular positions (top two traces) is supported by stable steps in firing rate by oculomotor integrator neurons (bottom two traces) that integrate transient ($\sim$ 100 ms) saccadic command bursts. Reproduced with permission from [140]. (b) Oculomotor neurons maintain eye position through linearly ramping tuning curves (bottom); responses are the same in the light and dark (top) and thus do not depend on visual input for gaze stabilization on the time-scale of seconds. Reproduced with permission from [270]. (c) Transient current injection into single oculomotor neurons reveals a transient, not persistent, decrease or elevation (inset) in firing rate, consistent with lack of a cellular origin for persistent intersaccadic firing. Reproduced with permission from [140]. (d) Injection of kainic acid into the oculomotor integrator produces leaky dynamics even in the light (inset, faster leak in the dark), consistent with network models. Adapted with permission from [271]. (e) Visual feedback mimicking leaky or unstable eye positions in goldfish can mistune the oculomotor integrator, making it unstable or leaky, respectively. Adapted with permission from [272, 145].

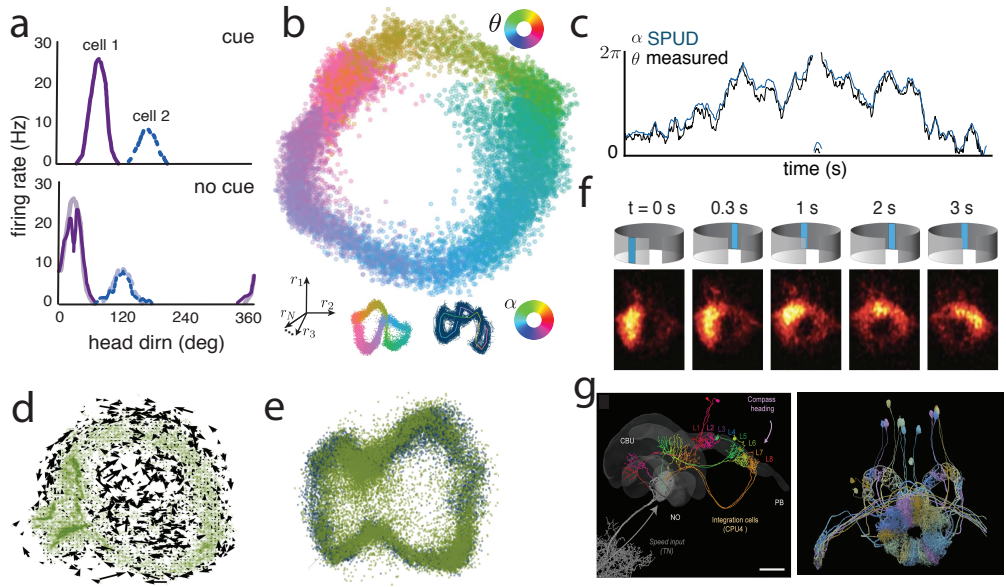

Figure 5: **The head direction circuit: a ring attractor in the brain.** (a) Activity of two cells in the rat HD circuit during free foraging in a 2-dimensional circular arena with a globally orienting cue (top). When the cue is removed (bottom), the fields rotate but the cells maintain their tuning shapes and relative tuning angles (pale curves: top plot, globally rotated. Adapted with permission from [75, 98]. (b) A nonlinear 2-dimensional embedding of the population-level states of the thousands-of-neurons sized mammalian thalamic area ADn recorded during 2-dimensional free-foraging: the states are confined to a 1-dimensional ring (cf. Fig. 1c); here, colors encode the measured head direction of the rodent. Inset: Non-linear embedding of states from a different rodent (left), with coloring obtained by an isometric parametrization along the ring by SPUD [19]. Reproduced with permission from [19]. (c) A close match between unsupervised isometric parametrization of the manifold from (b, inset) and the externally measured head direction of the rodent. Reproduced with permission from [19]. (d-e) The same cells as in (b, inset), recorded during REM sleep (green): the states during REM remain confined to a 1-dimensional ring that precisely overlays the ring of waking states (blue, in (e)), and exhibit large flows back toward the ring (in (d)) Reproduced with permission from [19]. (f) Calcium imaging of activity in the physically ring-shaped *Drosophila* ellipsoid body reveals a localized bump of excitation that follows the movement of a cue in the fly's visual field. Reproduced with permission from [156]. (g) Combination of electrophysiology and EM imaging of the central complex in bees (left) and flies (right) provides detailed layout and connectivity data for comparison with predicted connectivity in ring attractor models. Reproduced with permission from [161, 273].

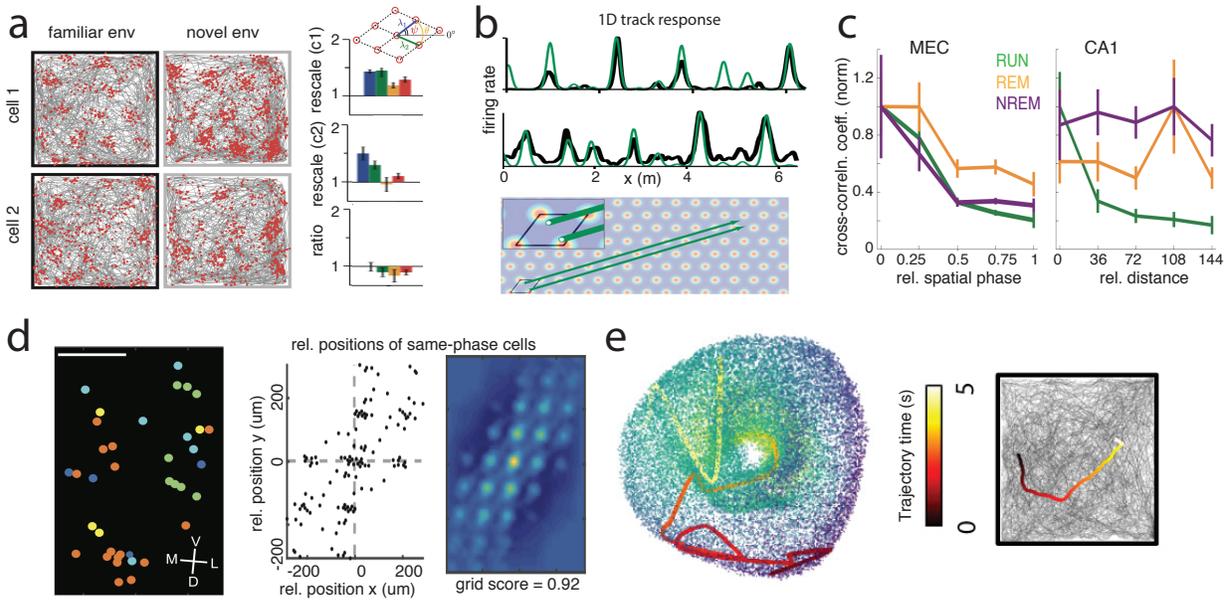

Figure 6: **A set of 2-dimensional toroidal attractors in the grid cell system.** (a) An example pair of grid cells (left column) whose spatial tuning periods and orientations reconfigure in novel environments (middle column), but the changes are tightly yoked to preserve cell-cell relationships (right column). Each color corresponds to a variable describing the lattice of the spatial tuning curve of the cell as shown in the schematic. Adapted with permission from [76]. (b) Responses of co-modular cells on 1D linear tracks can explained by parallel slices through a 2-dimensional grid, suggesting preserved 2-dimensional circuit dynamics across diverse environments. Reproduced with permission from [167]. (c) Pairwise correlations between grid cells measured during navigation are preserved across overnight sleep, while those of place cells are not. Reproduced with permission from [100, 101]. (d) Grid cells are anatomically arranged according to their relative spatial firing phases. (left) Cell positions are colored according to the phase of their spatial tuning curves. The relative cortical positions of same-phase cells make a triangular lattice pattern (middle), with a grid-like autocorrelation pattern (right). Reproduced with permission from [274] (e) Non-linear dimensionality reduction and topological data analysis directly reveal that the states of individual grid modules lie on a torus (left); as the animal follows a spatial trajectory (right), the state moves along the manifold (left). Reproduced with permission from [20].

# Acknowledgements

I.R.F. acknowledges funding from the Simons Foundation, ONR, HHMI through the Faculty Scholars Program, the Department of Brain and Cognitive Sciences, MIT, and the McGovern Institute, MIT. M.K. is supported by a Friends of the McGovern Institute Fellowship, a MathWorks Fellowship, and the Department of Physics, MIT.
We are grateful to Kayvon Daie, the anonymous reviewers, Natasha Bray, and Sarthak Chandra and other members of the Fiete lab for helpful comments on the manuscript.